\documentclass[twocolumn,prb,showpacs,floatfix]{revtex4}
\usepackage{amssymb}

\usepackage{amsmath}
\usepackage{epsfig}
\usepackage{graphics}

\begin{document}

\title{Universal spin-polarization fluctuations in 1D wires with magnetic
impurities}
\author{N.~A. Mortensen}
\email{nam@crystal-fibre.com}
\affiliation{Mikroelektronik Centret, Technical University of Denmark,DK-2800 Kgs.
Lyngby, Denmark}
\affiliation{{\O}rsted Laboratory, Universitetsparken 5, DK-2100 Copenhagen{\O},Denmark}
\author{J.~C. Egues}
\email{egues@if.sc.usp.br}
\affiliation{Departamento de F\'{\i}sica e Inform\'{a}tica, Instituto de F\'{\i}sica de S%
\~{a}o Carlos Universidade de S\~{a}o Paulo, 13560-970 S\~{a}o Carlos, S\~{a}%
o Paulo, Brazil}
\affiliation{Department of Physics and Astronomy, University of Basel, Klingelbergstrasse
82, CH-4056 Basel, Switzerland }

\begin{abstract}
We study conductance and spin-polarization fluctuations in one-dimensional
wires with spin-5/2 magnetic impurities (Mn). Our tight-binding Green
function approach goes beyond mean field thus including \textit{s-d}
exchange-induced spin-flip scattering. In a certain parameter range, we find
that spin flip suppresses conductance fluctuations while enhancing
spin-polarization fluctuations. More importantly, spin-polarization
fluctuations attain a \textit{universal value} $1/3$ for large enough spin
flip strengths. This intrinsic spin-polarization fluctuation may pose a
severe limiting factor to the realization of steady spin-polarized currents
in Mn-based 1D wires.
\end{abstract}

\date{\today}
\maketitle

\section{Introduction}

Spin-related effects in novel solid state heterostructures give
rise to a rich variety of fascinating physical phenomena. These
spin-dependent properties also underlie a potential technological
revolution in conventional electronics.\cite{wolf_etal_2001} This
new paradigm is termed ``Spintronics''. A particularly interesting
theme within this emerging field is spin-polarized transport in
semiconductor heterostructures. This topic has attracted much
attention after the fundamental discovery of exceedingly long spin
diffusion lengths in doped semiconductors \cite{kik99} followed by
the seminal spin injection experiments in Mn-based heterojunctions. \cite%
{spin_injection}

Theoretically, a number of works have addressed issues connected
with spin-polarized transport. These include, for instance: spin
filtering,\cite{spin-filtering} spin waves,\cite{farinas01} and quantum shot noise,\cite%
{brito2001} - all in ballistic semimagnetic tunnel junctions - and
mesoscopic conductance fluctuations in Rashba
wires.\cite{nikolic2001} Spin-dependent phenomena in connection
with localization effects should bring about exciting new physics.

Here we investigate conductance \textit{and} spin-polarization
fluctuations for transport through one-dimensional wires with
spin-5/2 magnetic impurities, \textit{e.g.}, Mn-based II-VI alloys
such as ZnSe/ZnMnSe/ZnSe. The experimental feasibility of these
wires has already been demonstrated.\cite{dietl,experiments} In
these systems, the conduction electrons interact with the
localized \emph{d} electrons of the Manganese via the \emph{s-d}
exchange coupling.\cite{furdyna1988} UCF in Mn-based submicron
wires was first experimentally studied in Ref.
[\onlinecite{dietl}]. We describe transport within the Landauer
formalism \cite{landauer} and calculate the relevant transmission
coefficients via non-interacting tight-binding Green
functions.\cite{datta}

\begin{figure*}[ht!]
\epsfig{file=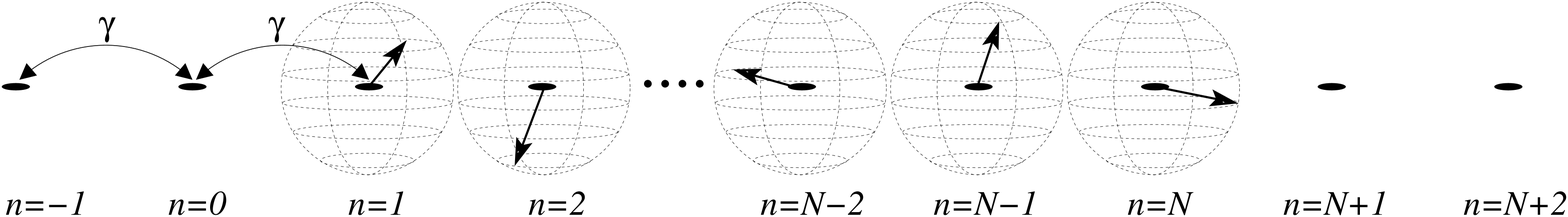, width=0.7\textwidth}
\caption{One-dimensional tight-binding chain with $N$ magnetic $s=5/2$
impurities (mutually un-correlated, each spin equally distributed among the
6 spin states) coupled to ideal impurity-free leads (sites $n<1$ and $n>N$).}
\label{fig:model}
\end{figure*}

We treat the \emph{s-d} interaction beyond the usual mean-field theory thus
accounting for spin flip scattering. In a certain parameter range we find
that spin-flip scattering suppresses conductance fluctuations\cite{dietl-exp}
(below the UCF value for strictly 1D wires) while enhancing the
corresponding spin-polarization fluctuations. More importantly, we show that
the spin-polarization fluctuations attain a \textit{universal value} $%
\langle (\delta \zeta )^{2}\rangle =1/3$ for strong spin-flip scattering.
This large spin-polarization fluctuation may pose a fundamental obstacle to
attaining steady spin-polarized currents in Mn-based wires.

\section{Model Hamiltonian}

We consider a one-dimensional tight-binding chain, see Fig.~%
\ref{fig:model}, of $N$ spin $s=5/2$ magnetic impurities coupled to ideal
leads (sites $n<1$ and $n>N$). We separate the electronic and impurity-spin
degrees of freedom and treat the latter classically (static scatterers). The
two-component electron wave function, $\psi =(\psi _{\uparrow },\psi
_{\downarrow })$ is then governed by the Schr\"{o}dinger equation with a
Hamiltonian
\begin{equation}  \label{Hamiltonian}
{\boldsymbol H}=%
\begin{pmatrix}
{\boldsymbol H}_0 & {\boldsymbol 0} \\
\boldsymbol{0} & {\boldsymbol H}_0%
\end{pmatrix}
+
\begin{pmatrix}
{\boldsymbol H}_{\uparrow\uparrow} & {\boldsymbol H}_{\uparrow\downarrow} \\
{\boldsymbol H}_{\downarrow\uparrow} & {\boldsymbol H}_{\downarrow\downarrow}%
\end{pmatrix}%
.
\end{equation}
Here, ${\boldsymbol H}_0$ is spin independent with elements \cite{datta}
\begin{equation}
\{{\boldsymbol H}_{0}\}_{nm}=2\gamma \delta _{nm}-\gamma \delta
_{nm+1}-\gamma \delta _{nm-1}+V_{n}\delta _{nm},
\end{equation}
where $V_{n}$ is the potential at site $n$ and $\gamma =\hbar ^{2}/2ma^{2}$,
with $a$ being the ``lattice constant''. In the leads ${\boldsymbol H}_{0}$
itself gives rise to the usual dispersion relation $\varepsilon (k)=2\gamma
(1-\cos ka)$.

In the following, $\sigma =\uparrow \;\equiv 1/2$ and $\sigma =\downarrow
\;\equiv -1/2$. We restrict ourselves to zero magnetic field so that the
block-matrices ${\boldsymbol H}_{\sigma \sigma }$ have elements given by
\begin{equation}
\{{\boldsymbol H}_{\sigma \sigma }\}_{nm}=\delta _{nm}J_{z}\sigma S_{n,z}
\label{longitudinal}
\end{equation}
which is a Heisenberg-like interaction of the spin of the electron ($\sigma $%
) with the $z$-component spin of the impurity $S=(S_{x},S_{y},S_{z})$. The
off-diagonal block-matrix ${\boldsymbol H}_{\uparrow \downarrow }={%
\boldsymbol H} _{\downarrow \uparrow }^{\dagger }$ contains the interaction
of the electron spin with the $x$ and $y$-components of the impurity spins
which leads to spin flip
\begin{equation}
\{{\boldsymbol H}_{\uparrow \downarrow }\}_{nm}=\delta _{nm} \big[%
J_{x}S_{n,x}-iJ_{y}S_{n,y}\big]/2.  \label{flip}
\end{equation}

\section{Transport properties}

We consider a sufficiently weak coupling between the impurity spins so that
they can be considered mutually uncorrelated, \emph{i.e.}\textit{,} no
magnetic ordering. The $z$-component of each spin is equally distributed
among the 6 spin states and the $x$ and $y$-components are uniformly
distributed with the constraint that $%
S^{2}=S_{x}^{2}+S_{y}^{2}+S_{z}^{2}=s(s+1)$, see Fig.~\ref{fig:model}.

We study transport in the low-temperature linear response limit
within the Landauer formalism\cite{landauer}
\begin{equation}
g =\frac{e^{2}}{h}\sum_{\sigma \sigma ^{\prime }} {\boldsymbol T}_{\sigma
\sigma ^{\prime }}(\varepsilon _{F}).  \label{landauer}
\end{equation}
Here, ${\boldsymbol T}$ is a $2\times 2$ matrix with the elements ${%
\boldsymbol T}_{\sigma \sigma ^{\prime }}$ being the transmission
probability of an electron from a state with spin $\sigma ^{\prime }$ in one
lead to a state with spin $\sigma $ in the other lead. From Eq.~(\ref%
{landauer}) we now define the degree of spin polarization
\begin{equation}
\zeta \equiv \frac{I_{\uparrow }-I_{\downarrow }}{I_{\uparrow}+I_{\downarrow
}} =\frac{{\boldsymbol T}_{\uparrow \uparrow } +{\boldsymbol T}_{\uparrow
\downarrow } -{\boldsymbol T}_{\downarrow \uparrow } -{\boldsymbol T}%
_{\downarrow \downarrow }} {{\boldsymbol T}_{\uparrow \uparrow } +{%
\boldsymbol T}_{\uparrow \downarrow } +{\boldsymbol T}_{\downarrow \uparrow
} +{\boldsymbol T}_{\downarrow \downarrow }},  \label{pol}
\end{equation}
which we will focus on in this paper.

\emph{Green function method.} The transmission matrix $\boldsymbol T$ is
related to the retarded Green function
\begin{equation}  \label{retarded}
{\boldsymbol G}(\varepsilon)=(\varepsilon\cdot{\boldsymbol 1} -\tilde{%
\boldsymbol H}-{\boldsymbol \Sigma}(\varepsilon))^{-1}
\end{equation}
via the Fisher--Lee relation \cite{fisher1981}
\begin{equation}
{\boldsymbol T}_{\sigma \sigma^{\prime}}(\varepsilon ) =[\hbar
v(\varepsilon)]^{2} \big|\{{\boldsymbol G}_{\sigma
\sigma^{\prime}}(\varepsilon )\}_{N1}\big|^{2},  \label{lee-f}
\end{equation}
where $v=\hbar^{-1} \partial \varepsilon /\partial k$ is the group velocity
in the leads. In Eq.~(\ref{retarded}) the $2N\times 2N$ matrix $\tilde{%
\boldsymbol H}$ is the Hamiltonian truncated to the $N$ lattice
sites with magnetic impurities. The effect of coupling to the
leads is contained in the $2N\times 2N$ retarded self-energy
matrix with elements\cite{datta}
\begin{equation}
\{{\boldsymbol\Sigma }_{\sigma \sigma ^{\prime }}(\varepsilon
)\}_{nm}=-\gamma e^{ik(\varepsilon )a}\delta _{\sigma \sigma ^{\prime
}}\delta _{nm}(\delta _{1n}+\delta _{Nn}).
\end{equation}

\emph{$N=1$ case.} A chain with a single impurity is a simple illustrative
example where analytical progress is possible. After performing the
straightforward matrix inversion in Eq.~(\ref{retarded}) we find
\begin{equation}
\zeta (\varepsilon )=\frac{-V_{1}J_{z}S_{z}} {V_{1}^{2}+\varepsilon
(4\gamma-\varepsilon )+(J\cdot S)^{2}}
\end{equation}
where $J=(J_{x},J_{y},J_{z})^{T}$. In zero magnetic field $\langle
S_{z}\rangle =0$ and $\big<S_{z}^{2}\big>=35/12$. This implies that $\langle
\zeta \rangle =0$ both with and without spin flip whereas the fluctuations
are finite. The analytical averaging is of course complicated by the
presence of $S_{z}$ in the denominator, but for isotropic coupling $%
J_{x}=J_{y}=J_{z}=J_{0}$, we have $(J\cdot S)^{2}=J_{0}^{2}s(s+1)$ so that $%
S_{z}$ only shows up in the numerator, \textit{i.e.}
\begin{equation}
\big<(\delta \zeta )^{2}\big>=\frac{35}{12}\frac{V_{1}^{2}J_{0}^{2}} {%
[V_{1}^{2}+\varepsilon (4\gamma -\varepsilon )+J_{0}^{2}s(s+1)]^{2}}.
\end{equation}
In the absence of spin-flip ($J_{x}=J_{y}=0$) the fluctuations are enhanced
due to the replacement of $s(s+1)\rightarrow S_{z}^{2}<s(s+1)$ in the
denominator (the final expression for the fluctuations is much more
complicated) and this means that spin-flip will lower the fluctuations of $%
\zeta $! Of course this trend is strictly valid for $N=1$, but in a limited
parameter range, this trend is still true for larger $N$ values.

\emph{Finite $N$ case.} For a finite number of impurities the problem is not
analytically tractable and we study the problem numerically by generating a
large ensemble (typically $10^{5}$ members) of spin configurations. For each
spin configuration we calculate Eqs.~(\ref{retarded},\ref{lee-f})
numerically. In our simulations we use the following parameters: $%
\varepsilon _{F}=\gamma $, $J_{z}=\gamma /2$, $V_{n}=0$ (\textit{i.e.}, we
neglect \emph{spatial disorder}), and varying spin-flip coupling strengths $%
0\leqslant J_{x}=J_{y}\leqslant \gamma $.

\begin{figure}[th]
\begin{center}
\epsfig{file=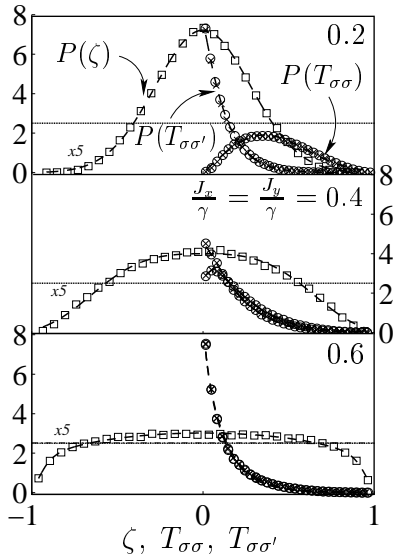, width=0.3\textwidth}
\end{center}
\par
\vspace{-5mm}
\caption{Distributions $P(\protect\zeta )$, $P(T_{\protect\sigma \protect%
\sigma })$, and $P(T_{\protect\sigma {\protect\sigma}^{\prime}})$ for
different spin-flip scattering strengths $J_{x}=J_{y}$ in the case of $N=10$%
. The dash-dotted line in the lowest panel indicates the uniform limit $P(
\protect\zeta )=1/2$ [note the magnification of $P(\protect\zeta)$ by of
factor of $5$] attained for strong enough spin-flip scattering.}
\label{fig:P}
\end{figure}

\section{Results and discussions}

\begin{figure}[th]
\begin{center}
\epsfig{file=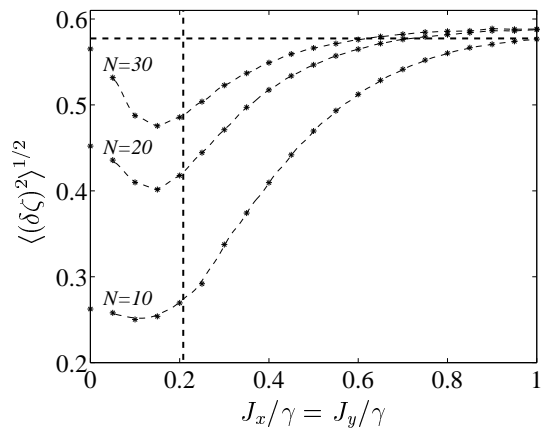, width=0.40\textwidth}
\end{center}
\par
\vspace{-5mm}
\caption{Average fluctuations $\big<(\protect\delta \protect\zeta )^{2}\big>%
^{1/2}$ as a function of spin-flip strength for $N=10$, $20$, and $30$. The
dashed horizontal line indicates the universal value $1/\protect\sqrt{3}$
obtained from the uniform limit $P(\protect\zeta )=1/2$. The vertical dashed
line indicates where the spin-flip rate is comparable to $|J_{z}\protect%
\sigma |/\hbar $.}
\label{fig:fluc_zeta}
\end{figure}

Figure~\ref{fig:P} shows the distributions $P(\zeta )$, $P(T_{\sigma \sigma
})$, and $P(T_{\sigma \sigma^{\prime}})$ for $N=10$ and increasing strengths
of the spin-flip coupling $J_{x}=J_{y}$. The distribution $P(\zeta )$ is
symmetric around $\zeta =0$ which implies that on average there is no spin
filtering, $\langle \zeta \rangle $ $=0$. The distribution $P(\zeta )$ first
gets narrower for spin flip in the $[0,0.15\gamma ]$ range (not shown) and
then broadens as spin flip further increases. For sufficiently strong
spin-flip scattering the distribution approaches that of the uniform limit
in which $P(\zeta )=1/2$. In this limit $P(T_{\sigma \sigma })$ and $%
P(T_{\sigma \sigma^{\prime}})$ coincide and so do all average transmission
probabilities $\langle {T}_{\sigma \sigma ^{\prime }}\rangle .$ As we
discuss below, the initial narrowing and subsequent broadening of $P(\zeta )$
with spin flip gives rise to a minimum in the fluctuation of $\zeta $, Fig.~%
\ref{fig:fluc_zeta}.

\emph{Universal spin-polarization fluctuations.} In the limit of a short
spin-flip length $\ell _{\sigma }\ll L$ we in general find a uniform
distribution $P(\zeta )=1/2$, Fig.~\ref{fig:P}. This uniform distribution
yields the universal value $\langle (\delta \zeta )^{2}\rangle =1/3$ for the
spin-polarization fluctuations. Figure~\ref{fig:fluc_zeta} clearly shows
that this universal value is attained for increasing spin flip strengths and
is indeed independent of $N$. Interestingly, Fig.~\ref{fig:fluc_zeta} also
shows a minimum at around $J_{x}=J_{y}=0.15\gamma $. This minimum can be
attributed to two competing energy scales:\ the longitudinal $(\sim J_{z})$
and the transverse $(\sim J_{x},J_{y})$ parts of the \emph{s-d} exchange
interaction, Eqs. (\ref{longitudinal}) and (\ref{flip}), respectively. A
simple ``back-of-the-envelope'' calculation shows that these two competing
scales are equal for $J_{x}=J_{y}=J_{z}/\sqrt{2s(s+1)/3}\gamma =0.208\gamma $%
. The vertical dashed line in Fig.~\ref{fig:fluc_zeta} indicates this value.
Observe that $\langle (\delta \zeta )^{2}\rangle $ becomes larger for
increasing $N$. This happens because $P(\zeta )$ broadens for larger $N$'s
(the traversing electrons see a wider region with random spins).\ This is
similar to the broadening due to increasing spin flip strength.

We should mention that the distribution $P(\zeta )$, and consequently $%
\langle (\delta \zeta )^{2}\rangle $, change dramatically for $\varepsilon
_{F}<J_{z}$. In this regime, $P(\zeta )$ becomes \emph{U}-shaped (not shown)
because of the dominant filtering due to the ``end states'' with $%
S_{j,z}=\pm 5/2$. This \emph{qualitatively} different $P(\zeta )$ yields a
monotonically decreasing $\langle (\delta \zeta )^{2}\rangle $ as a function
of spin-flip strength. Here the universal $\langle (\delta \zeta
)^{2}\rangle =1/3 $ value is approached from above for large spin flip
strengths ($J_{x}=J_{y}\sim\gamma $).

\begin{figure}[th]
\begin{center}
\epsfig{file=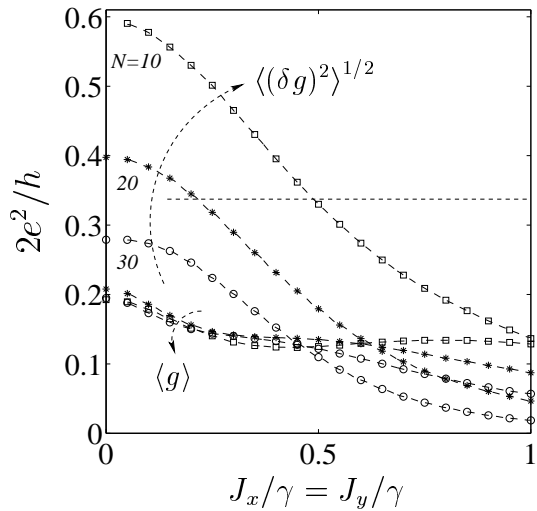, width=0.40\textwidth}
\end{center}
\par
\vspace{-5mm}
\caption{Average conductance $\langle g \rangle$ and its fluctuations $\big<(%
\protect\delta g)^{2}\big>^{1/2}$ as a function of the spin-flip scattering
strength for $N=10$, $20$, and $30$. The conductance fluctuations are much
more sensitive to spin flip than the average conductance: the former gets
strongly suppressed for increasing spin flip rates.}
\label{fig:g}
\end{figure}

\emph{Suppression of conductance fluctuations.} Whereas the fluctuations in
the spin polarization $\zeta $ remain finite in the strong spin-flip
scattering regime (Fig.~\ref{fig:fluc_zeta}), we find that the fluctuations
of the conductance $g$ are strongly suppressed in this limit. This is
illustrated in Fig.~\ref{fig:g} which shows the average conductance and its
fluctuations as a function of spin-flip scattering for $N=10$, $20$, and $30$%
. Note that $\langle (\delta g)^{2}\rangle ^{1/2}$ is much more sensitive to
spin flip than $\langle g\rangle $. In addition, for all $N$ we essentially
have $\langle (\delta g)^{2}\rangle ^{1/2}>\langle g\rangle $ for $%
J_{x}=J_{y}\rightarrow 0$ and $\langle (\delta g)^{2}\rangle ^{1/2}\lesssim
\langle g\rangle $ for $J_{x}=J_{y}\rightarrow \gamma $. Figure \ref{fig:g}
clearly shows the conductance fluctuations get suppressed for increasing $N$%
. The horizontal dashed line shows the UCF\ value ($0.73/2=0.365$, see
\textit{e.g.} Ref.~\onlinecite{UCF}) for a 1D\ wire in the metallic regime.
The spin-related \emph{conductance} fluctuations do not approach a finite
value for increasing spin flip scattering. It actually seems to go to zero.
This is in contrast to the \emph{spin-polarization} fluctuations, Fig.~\ref%
{fig:fluc_zeta}, which attain a universal value $\langle (\delta \xi
)^{2}\rangle ^{1/2}=1/\sqrt{3}$ for strong spin-flip scattering.
Incidentally, we observe that $\langle (\delta g)^{2}\rangle ^{1/2}$ and $%
\langle (\delta \zeta )^{2}\rangle ^{1/2}$ also present contrasting behavior
for increasing $N$ (and $\varepsilon _{F}>J_{z}$): the former gets
suppressed while the latter gets enhanced (cf. Figs. \ref{fig:fluc_zeta} and %
\ref{fig:g}).

\emph{Spin disorder as spatial disorder.} To some extent, the \emph{s-d}
site interaction considered here plays the role of spatial disorder in the
system with a mean-free path $\ell_J$. Let us consider first the case with
no spin flip (\textit{i.e.}, $J_{x}=J_{y}=0$). In this case, the term $%
J_{z}\sigma S_{n,z}$ acts as a ``random'' spin-dependent potential along the
chain (here the site potential has some internal structure). As shown in
Fig.~\ref{fig:g} the conductance fluctuations for zero spin-flip scattering
are larger than, slightly above, and slightly below, the UCF value for $N=10$%
, $N=20$, and $N=30$, respectively. For increasing $N$ we go from
the metallic regime ($L=Na\ll \ell_{J_z}$) with vanishing
fluctuations and a Gaussian $P(g)$ strongly peaked near $g\sim
2e^2/h$ to the strongly localized regime ($L\gg \ell_{J_z}$) where
it is well-known that $P(g)$ is strongly peaked near $g\sim 0$
with a log-normal distribution so that fluctuations can be
comparable to the mean value.\cite{beenakker1997} This is in
accordance with numerical studies with different continuous
distributions of the ``on-site'' potential (\textit{e.g.} Gaussian
or uniform distributions).\cite{kramer1993} In Fig.~4 the
``small'' mean values $\langle g \rangle$, for $N=10,20$, and 30,
indicate the onset of localization with fluctuations comparable to
the mean value. As $N$ gets
larger conductance fluctuations are as expected suppressed \cite%
{beenakker1997,mirlin2000}.

\emph{Role of spin-flip scattering}. Spin flip clearly suppresses
conductance fluctuations, Fig.~\ref{fig:g}. This can be understood from
Eq.~(4) being a \textit{complex} number with a random phase which makes spin
flip act as a source of ``de-coherence'' [the total wave function is, of
course, fully coherent]. Furthermore, spin flip mixes all the $S_{n,z}$
components on each site thus smoothing the potential seen by the traversing
electron and hence reducing conductance fluctuations.\ This is true for both
$\varepsilon _{F}>J_{z}$ [except for the window $[0,0.15\gamma ]$ in which $%
P(\zeta )$ narrows] and $\varepsilon _{F}<J_{z}$.

\emph{``Truly'' universal fluctuations.} Why is $\langle (\delta \zeta
)^{2}\rangle ^{1/2}$ universal even for short spin-flip lengths $\ell
_{\sigma }\ll L$ (strong spin-flip scattering) while $\langle (\delta
g)^{2}\rangle ^{1/2}$ is clearly suppressed below the usual UCF value in
this limit? It is well known that conductance fluctuations are suppressed in
the incoherent limit. \cite{UCF} More specifically, in 1D wires with $%
\ell_{\varphi }\ll L$, $\ell_{\varphi }$ is some ``dephasing
length'', the suppression factor is $\sqrt{L/\ell_{\varphi }}$
(see Ref. [\onlinecite{datta}]). Interestingly, we can likewise
understand the suppression of $\langle (\delta g)^{2}\rangle
^{1/2}$ seen in our simulations by viewing spin-flip scattering as
producing ``dephasing'' with $\ell_{\varphi }\sim \ell_{J_{x,y}}$
\cite{dephasing}. For the spin-polarization fluctuations,
however, the picture is slightly different: here we divide our system in $%
N_{L}=L/l_{\varphi }$\emph{\ }segments. To each of these we can associate an
average\ spin polarization $\langle \zeta _{i}\rangle =0$ $(i$: $1..N_{L})$
\ and a corresponding spin-polarization fluctuation $\langle (\delta \zeta
_{i})^{2}\rangle $. Neither $\langle \zeta _{i}\rangle $ nor $\langle
(\delta \zeta _{i})^{2}\rangle $ are additive quantities like $\langle
g\rangle $ and $\langle (\delta g)^{2}\rangle $ (``extensive versus
intensive'' properties). Sensible \emph{global} averages for the whole
system are then $\bar{\zeta}\equiv \frac{1}{N_{L}}\sum_{i}\langle \zeta
_{i}\rangle =0$ and $\overline{{(\delta \zeta )}^{2}}\equiv \frac{1}{N_{L}}%
\sum_{i}\langle (\delta \zeta _{i})^{2}\rangle $. We should expect $%
\overline{{(\delta \zeta )}^{2}}=$ $\langle (\delta \zeta _{i})^{2}\rangle
\equiv \langle (\delta \zeta )^{2}\rangle $ if the system is \emph{ergodic}.
Hence universal spin-polarization fluctuations are not suppressed for large
spin-flip scattering in contrast to conductance fluctuations.

\section{Concluding remarks}

Spin-flip scattering in Mn-based wires reduces conductance
fluctuations while enhancing spin-polarization fluctuations in a
limited parameter range. Remarkably, spin-polarization
fluctuations reach a universal value $1/3$ for large spin-flip
scattering in which the conductance fluctuations vanish. This
universal value should manifest itself in time- and
polarization-resolved photo-luminescence measurements. More
important, these sizable spin fluctuations may limit the
possibilities for steady spin injection in these systems.

The authors thank A.-P. Jauho for a critical reading of the manuscript. We
also acknowledge K. Flensberg (NAM), U. Z\"{u}licke and M. Governale (JCE)
for useful discussions. This work has been supported by the Swiss NSF,
DARPA, ARO, and FAPESP/Brazil (JCE).


\begin{thebibliography}{99}
\bibitem{dephasing} Again, this is a sensible view since spin flip is a
complex number with a random phase in our model, see Eq. (\ref{flip}).
\end{thebibliography}
\end{document}